\documentclass[
    nobibnotes, aps, preprint,
    superscriptaddress,
    obeyspaces, spaces,
]{revtex4-2}

\usepackage[english,main=english]{babel}

\usepackage{multirow}
\usepackage{xcolor}

\usepackage{lmodern}                
\usepackage[T1]{fontenc}            
\usepackage[utf8]{inputenc}         
\usepackage[english,main=english]{babel}

\usepackage{graphicx}               
\usepackage{longtable}
\usepackage{makecell}               
\usepackage{booktabs}               
\usepackage{dcolumn}                
\usepackage{setspace}               
\usepackage[
    colorlinks,
    citecolor=blue,
    urlcolor=blue,
    filecolor=blue,
    linkcolor=black
]{hyperref}
\usepackage{ragged2e}

\usepackage{soul}                   
\usepackage{color}                  
\usepackage{xspace}                 
\usepackage[mathlines]{lineno}      

\begin{document}

\title[]{
    Benchmarking Large Language Models for Polymer Property Predictions
}

\author{Sonakshi Gupta}
    \affiliation{
        School of Computational Science and Engineering,
        Georgia Institute of Technology, 771 Ferst Drive NW, Atlanta 30332,
        GA, USA.
    }
\author{Akhlak Mahmood}
    \affiliation{
        School of Materials Science and Engineering,
        Georgia Institute of Technology, 771 Ferst Drive NW, Atlanta 30332,
        GA, USA.
    }
\author{Shivank Shukla}
    \affiliation{
        School of Materials Science and Engineering,
        Georgia Institute of Technology, 771 Ferst Drive NW, Atlanta 30332,
        GA, USA.
    }
\author{Rampi Ramprasad}
    \email{rampi.ramprasad@mse.gatech.edu}
    \affiliation{
        School of Materials Science and Engineering,
        Georgia Institute of Technology, 771 Ferst Drive NW, Atlanta 30332,
        GA, USA.
    }
    \altaffiliation[Email: ]{\url{rampi.ramprasad@mse.gatech.edu}}
    \email{rampi.ramprasad@mse.gatech.edu.}


\begin{abstract}
Machine learning and artificial intelligence have revolutionized polymer science by enhancing the ability to rapidly predict key polymer properties and enabling generative design.
The utilization of large language models (LLMs) in polymer informatics may offer additional opportunities for advancement. Unlike traditional methods that depend on large labeled datasets, hand-crafted representations of the materials, and complex feature engineering, LLM-based methods utilize natural language inputs via a transfer learning process and eliminate the need for complex representation and fingerprinting, thus significantly simplifying the training process.
In this study, we fine-tune general-purpose LLMs—open-source LLaMA-3-8B and commercial GPT-3.5—on a curated dataset of 11,740 entries to predict key thermal properties: glass transition ($T_g$), melting ($T_m$), and decomposition ($T_d$) temperatures. Using parameter-efficient fine-tuning and hyperparameter optimization, we benchmark these models against traditional fingerprinting-based approaches including Polymer Genome, polyGNN, and polyBERT, under both single-task (ST) and multi-task (MT) learning frameworks.
We find that while LLM informatics techniques can come close to traditional methods, they generally underperform in terms of predictive accuracy and computational efficiency.
The fine-tuned LLaMA-3 model consistently outperforms GPT-3.5, likely due to the flexibility and tunability of the open-source architecture.
Additionally, ST learning proves more effective than MT for LLMs, which struggle to exploit cross-property correlations—a significant and known advantage of traditional methods.
The analysis of molecular embeddings learned by the models provides insight into the inner workings of the LLMs, revealing fundamental limitations of general-purpose LLMs in capturing nuanced chemo-structural information compared to the handcrafted features and domain-specific embeddings utilized in the traditional methods.
These findings offer insights into the interplay between molecular embeddings and natural language processing, and provide guidance for LLM model selection within the context of polymer informatics.
\end{abstract}


\maketitle
\doublespacing
\raggedbottom

\section*{Main}
Machine learning (ML) techniques are beginning to favorably impact materials property predictions and design \cite{ramprasad2017machine, tran_design_2024}. ML-assisted efforts are contributing to the design of capacitive energy storage systems \cite{gurnani_ai-assisted_2024, kern_design_2021, gurnani_polyg2g_2021}, fuel cell materials \cite{schertzer2025ai, wang2023large}, membranes for the separation of mixture of gases and solvents \cite{lee2023data, wang2023machine, phan2024gas}, recyclable polymers \cite{kern2024informaticsframeworkdesignsustainable, atasi2024design, guarda2024machine} and so on. 

Traditional ML approaches in polymer informatics typically follow a two-step process: first, transforming polymer structures into numerical representations known as fingerprints; second, applying supervised learning to predict target properties based on these representations. Over the years, numerous studies have been developed to enhance predictive performance. For instance, the hand-crafted Polymer Genome (PG) fingerprints represent polymers at three hierarchical levels—atomic, block, and chain—capturing structural details across multiple length scales \cite{kim2018polymer}. Graph-based methods like polyGNN employ molecular graphs to learn polymer embeddings, effectively balancing prediction speed and accuracy \cite{gurnani_polymer_2023, zang2023hierarchical, xia2022pretraining, wang2022molecular}. Transformer-based models such as polyBERT utilize the linguistic structure of SMILES (Simplified Molecular-Input Line-Entry System) strings~\cite{weininger1988smiles}, using an adaptation for polymers in which asterisks (\texttt{*}) mark the endpoints of repeating units, allowing for precise encoding of polymer structures \cite{kuenneth_polybert_2023,wang2019smiles, ross2022large, li2020learn}.


Once the materials' structures are fingerprinted, in the next stage, these representations are used to train downstream models in a supervised fashion that learn the relationships between structure and properties. Supervised learning algorithms can range from simple linear regression models to complex deep learning algorithms, such as Gaussian process regression (GPR), artificial neural networks (NNs), and graph neural networks (GNNs), tailored to specific properties and applications to ensure target performance and predictive accuracy. Furthermore, the performance of ML methods is often enhanced by multitask learning (MT), which uses correlations between multiple properties, allowing the model to learn them simultaneously even when smaller amount of data is available \cite{kuenneth_polymer_2021}.

Recently, LLMs have emerged as a promising tool for a range of tasks in materials science - from extracting information from literature \cite{gupta2024data}, predicting polymer solubility in solvents \cite{agarwal2025polymer}, to generating crystal structures \cite{gan2025large, antunes2024crystal}. LLM-generated embeddings have also demonstrated strong performance on high-dimensional regression tasks \cite{tang_understanding_2024}, providing a promising pathway to improve the predictive accuracy for molecular properties. When fine-tuned on materials-specific datasets, the models can interpret SMILES strings and predict properties directly from text, eliminating the need for handcrafted or graph-based fingerprints while offering scalability and simplicity. Recent work has shown substantial advances in molecular property prediction using LLMs \cite{liu_moleculargpt_2024, liu2023molxptwrappingmoleculestext, edwards2022translationmoleculesnaturallanguage, pei2024biot5generalizedbiologicalunderstanding, liu2024moleculargptopenlargelanguage}.

In polymer informatics, however, the potential of LLMs for property prediction remains largely unexplored. Various challenges arise because of the large size, repeating units, and structural complexity of polymers, which differ significantly from those of small molecules. Additionally, polymer property data are scarce compared to the available molecular databases, potentially limiting the ability of models to generalize across diverse polymer chemistries. Even for widely reported thermal properties, such as glass transition temperature ($T_g$), melting temperature ($T_m$), and decomposition temperature ($T_d$) -  the available datasets are only a fraction of what is available for molecules to build reliable and generalizable predictive models.
As a result, the predictive performance, strengths and limitations of LLMs compared to traditional ML approaches in polymer informatics are not fully understood.

In this work, we investigate the use of LLMs for polymer property prediction by fine-tuning Meta AI’s open-source LLaMA-3-8B-Instruct \cite{grattafiori2024llama3herdmodels} and OpenAI’s closed-source GPT-3.5 and GPT-4 models to predict key thermal properties - $T_g$, $T_m$, and $T_d$. We show that the general purpose, pre-trained, LLMs can be fine-tuned to predict polymer properties directly from the SMILES strings, eliminating the need for handcrafted or graph-based fingerprints.
We evaluate LLMs under three training strategies: single-task, multi-task, and continual learning, and perform a comprehensive comparison against traditional machine learning (ML) baselines to assess their respective strengths and limitations in the context of polymer informatics.
Our results show that fine-tuned LLMs achieve predictive performance that approach (but do not surpass) classical ML methods, with the LLaMA-3 models generally outperforming GPT-3.5—largely due to the flexibility of open-source fine-tuning and more effective hyperparameter optimization. Notably, we find that LLMs can jointly learn both polymer embeddings and structure–property relationships during fine-tuning, removing the need for a separate fingerprinting stage.
To evaluate the quality of these learned embeddings, we compare LLM-derived representations with domain-specific alternatives, including polyBERT, polyGNN, and Polymer Genome, assessing their effectiveness in capturing polymer structure–property relationships. Through this comparative study, we provide critical insights into the feasibility, advantages, and limitations of LLMs for polymer property prediction tasks, highlighting their potential as scalable alternatives to traditional ML models.

\begin{figure*}
    \centering
    \includegraphics[width=6.48in]{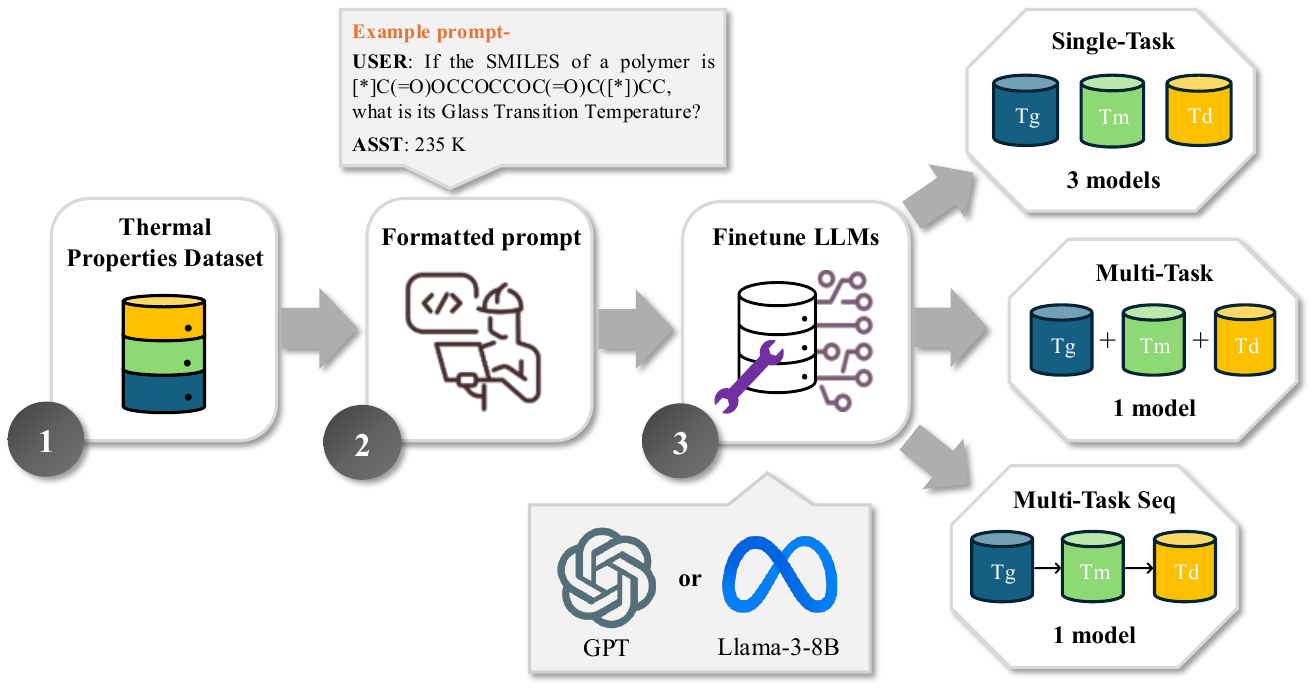}
    \caption{Overall workflow for adapting LLMs to predict thermal properties of polymers. (1) The thermal properties dataset, including Glass Transition Temperature (\(T_g\)), Melting Temperature (\(T_m\)), and Thermal Decomposition Temperature (\(T_d\)), is curated and processed into a standardized format using SMILES. (2) The dataset is transformed into an instruction-tuning format with an optimized prompt for property prediction tasks. (3) Large language models (LLaMA-3 and GPT-3.5) are fine-tuned using three strategies: Single-Task (ST), where separate models are trained independently for each property; Multi-Task (MT), where a single model is simultaneously trained on all three properties; and Multi-Task Sequential (MT-Seq), where properties are learned in a stepwise manner (\(T_g \rightarrow T_m \rightarrow T_d \)).}

\end{figure*}

\section*{RESULTS AND DISCUSSION}

\subsection*{Overview of the Property Prediction Pipeline}
In the polymer literature, thermal properties are the most frequently reported properties of homopolymers.
To train and assess the performance of LLMs for property prediction, we manually curated a benchmark dataset of experimental thermal property values containing the SMILES string and property values. The curated dataset contains $5,253$ Glass Transition Temperature (\(T_g\)), $2,171$ Melting Temperature (\(T_m\)), and $4,316$ Thermal Decomposition Temperature (\(T_d\)) values, as summarized in \textbf{Table 1}.

\begin{table}[!hbt]
\setlength{\tabcolsep}{6pt}
\centering
\caption{Benchmark thermal property datasets used to train the ML models and finetune the LLMs.}

\begin{tabular}{lcr}
\toprule
Property                            & \makecell[c]{Range (K)}   & \makecell[c]{Number of data points} \\
\midrule
Glass transition temperature	    & $[80.0, 873.0]$	  & $5,253$  \\
Melting temperature	                & $[226.0, 860.0]$	  & $2,171$  \\
Thermal decomposition temperature	& $[291.0, 1167.0]$	  & $4,316$  \\

\midrule
\textbf{Total}                      &  & $11,740$ \\

\bottomrule
\end{tabular}

\label{tab:proplist}
\end{table}

Polymer structures in the dataset are represented using SMILES, which provides a detailed and machine-readable format for polymer representations. However, a given polymer can have multiple syntactic variants of the SMILES string. To address this non-uniqueness, we performed canonicalization of SMILES to ensure a standardized and consistent representation for each unique polymer structure.

The dataset was subsequently transformed into an instruction-tuning format for fine-tuning LLMs. Since prompt design critically influences the performance of LLMs, a systematic prompt optimization process was employed to determine the most effective structure. The final prompt, which demonstrated superior accuracy, was structured as follows: \begin{quote} \textbf{User}: \texttt{If the SMILES of a polymer is <SMILES>, what is its <property>?}\
\textbf{Assistant}: \texttt{{smiles: <SMILES>, <property>: <value> <unit>}} \end{quote}
where the \texttt{<SMILES>}, \texttt{<property>}, \texttt{<value>} and \texttt{<unit>} placeholders were replaced by the actual SMILES string, property name, value and unit from the dataset for each row. The additional variations of the different prompts evaluated are listed in the Supporting Information and the corresponding performance are plotted in \textbf{Fig. S1}.

Model availability and architecture play a significant role in the performance of the LLMs. Open-source models such as Meta AI's LLaMA-3 provide greater flexibility for customization and control during fine-tuning, but require significant computational resources available on premise. OpenAI's GPT models, on the other end, offer ease of training and prediction with the utilization of Application Programming Interfaces (APIs), but are constrained by limited control over the hyperparameters one can tune and the black-box nature of the employed finetuning method. Understanding these trade-offs is essential for determining the most suitable approach for different polymer informatics tasks, balancing customization, computational efficiency, and ease of use.
We therefore fine-tuned both LLaMA-3 and GPT-3.5 models using the instruction-formatted dataset. For LLaMA-3, fine-tuning was carried out on in-house servers utilizing Low-Rank Adaptation (LoRA)~\cite{hu2021loralowrankadaptationlarge}, a parameter-efficient method that approximates large pre-trained weight matrices with smaller, trainable matrices. This approach significantly reduced computational overhead, accelerated training times, and lowered memory usage while preserving model performance. We optimized a range of hyperparameters, including the rank (\(r\)), scaling factor (\(\alpha\)), number of epochs, and softmax temperature ($T$) during inference to achieve the best results. In contrast, the fine-tuning of GPT-3.5 models was performed using the OpenAI API. Due to limited control over the fine-tuning process, the optimization of GPT-3.5 models was restricted only to the number of training epochs and the inference temperature, $T$. Additionally, the exact mechanism of the fine-tuning technique and model architecture implemented by OpenAI is not publicly available.

To comprehensively evaluate the performance of the models, we implemented three learning strategies: Single-Task (ST), Multi-Task (MT), and Multi-Task Sequential (MT-Seq). In the Single-Task framework, separate models were independently fine-tuned for each property, (\(T_g\), \(T_m\), and \(T_d\)), focusing exclusively on a single target at a time. The MT framework, by contrast, involved simultaneously fine-tuning a single model on all three properties, allowing the model to leverage shared parameters and capture inherent correlations across the properties.
By leveraging the correlations between properties, multitask framework reduces the risk of overfitting to any single property and improve predictive accuracy across all properties. This approach is supported by prior ML-based studies with deep neural networks \cite{kuenneth_polymer_2021} and graph neural network models \cite{gurnani_polymer_2023}.
The MT-Seq approach, inspired by the continual or transfer learning approach, extended this idea by fine-tuning a single model in a stepwise manner, beginning with \(T_g\), then \(T_m\), and finally \(T_d\), aiming to build on the knowledge gained from earlier tasks to improve adaptation and performance on the subsequent ones. A continual fine-tuning could enable progressive adaptation to later tasks hopefully without "forgetting" earlier ones, thus assessing the model’s ability to retain and build upon previously acquired knowledge. Our overall finetuning workflow is illustrated in \textbf{Fig.1}.

\subsection*{Optimization of Hyperparameters}

\subsubsection*{GPT Optimization}
OpenAI's GPT-4o was initially tested by fine-tuning on the \(T_g\) dataset using the ST method. However, the performance of GPT-4o was comparable to that of GPT-3.5 (\textbf{Fig. S2}), with no notable improvement in predictive accuracy.
We quantified the performance of the fine-tuned model by calculating the RMSE of the predictions made on the held-out test split of the dataset.
Given this similarity and performance of GPT-4o, we opted to proceed with GPT-3.5 for subsequent experiments due to its cost-effectiveness.
We optimized GPT-3.5 by testing different numbers of epochs (5 and 10) and the softmax temperatures (\(T = 0.5\) and \(T = 0.8\)) during the inference stage. Parity plots for all fine-tuned models, corresponding to \(T_g\), \(T_m\), and \(T_d\), are available in the Supplementary Information.
The best performance was achieved when GPT-3.5 was fine-tuned for 5 epochs and inference was conducted with \(T = 0.5\). Using these optimized hyperparameters, we also trained a multitask thermal property model (GPT-MT) and a multitask sequential model (GPT-MT-Seq) on the three thermal properties.
The final results of the fine-tuned GPT-3.5 models are detailed in the Performance section, where they are compared with LLaMA models across all tasks and training strategies.

\subsubsection*{LLaMa-3 Optimization}

The initial fine-tuning of the LLaMA-3-8B-Instruct model involved optimizing multiple additional hyperparameters, namely, alpha (\(a\)) and rank (\(r\)), in addition to evaluating the softmax temperatures (T) of 0.5 and 0.8 during inference. When T = 0.8, the combination of \(a = 16\) and \(r = 16\) yielded the best performance for the \(T_g\) dataset (cf. Fig. S2). Increasing both alpha and rank progressively to 128 resulted in a performance decline. This configuration (\(a = 16, r = 16, T = 0.8\)) also performed consistently well across the \(T_m\) and \(T_d\) datasets.
For \(T_m\), similar analysis revealed two optimum configurations, \((16, 16)\) and \((32, 32)\), that achieved comparable performance. Similarly, for \(T_d\), multiple configurations, including \((16, 16)\), \((32, 16)\), and \((64, 16)\), delivered similar results. It is important to note that stochastic variations across multiple inference trials can lead to slight fluctuations in these values, as elaborated in the later sections. Despite this variability, the results indicate a consistent pattern, with \((16, 16)\) emerging as the most effective hyperparameter combination for these datasets.
Building on our initial hyperparameters tuning, we explored the effect of the number of epochs, increasing them from 5 to 30 in increments of 5 while keeping \(a = 16\) and \(r = 16\). As shown in \textbf{Fig. S5a}, RMSE initially decreased with increasing epochs, reaching an optimal value at 25 epochs. At this point, the RMSE for the \(T_g\), \(T_m\), and \(T_d\) datasets at \(T = 0.8\) was 39.48~K, 56.89~K, and 75.79~K, respectively, with inference temperature \(T = 0.8\) consistently outperforming \(T = 0.5\). Interestingly, a similar RMSE minimum was observed at 5 epochs; however, upon closer inspection of the parity plots (cf. \textbf{Fig. S3}), significant clustering of the predicted values became evident, suggesting poor generalization and learning by the model despite the lower RMSE values.

This clustering phenomenon highlighted the need for an additional metric to better evaluate model performance.
From our experiments, similar clustering effects were observed for both LlaMa-3 and GPT-3.5 models (see \textbf{Fig. S4}), usually prominent at lower softmax temperatures.
We introduced the Bin Height Dispersion Index (BHDI), described in Methods section, to measure clustering in predictions. As shown in \textbf{Fig. S5b}, BHDI decreased with increasing epochs suggesting the absence of clustering, reaching its lowest at 25 epochs. For \(T_g\), BHDI dropped from 4.67 at 5 epochs to 0.98 at 25 epochs. Similarly, for \(T_m\), it decreased from 7.52 to 1.48, and for \(T_d\), from 7.03 to 2.02. The reduction in BHDI demonstrates that 25 epochs not only minimize clustering but also promote a well-distributed range of predictions, reflecting improved model performance. These results also underscored the importance of moving beyond traditional metrics like RMSE to evaluate the distribution, variability, and generalization of predictions.


\subsection*{Single-Task \textit{vs.} Multi-Task Performance}

The best performing fine-tuned LLM models were benchmarked against state-of-the-art (SOTA) approaches, including Polymer Genome (PG), polyGNN, and polyBERT, which primarily function as fingerprinting techniques to encode polymer structures for machine learning models. The training details for the SOTA approaches are provided in the Supporting Information. 

Among the traditional informatics methods, the PG-fingerprinting approach achieved the best performance, followed closely by polyGNN. Fine-tuned LLaMA-3 performed on par with polyBERT, a domain-specific chemical language model, demonstrating that a general-purpose LLM can achieve competitive accuracy without requiring extensive polymer-specific pretraining. However, LLMs did not surpass traditional ML models, reinforcing the advantage of domain-specific fingerprinting and feature engineering during polymer property prediction. \textbf{Fig. 2} presents the corresponding parity plots.

\begin{figure*}
    \centering
    \includegraphics[width=6.48in]{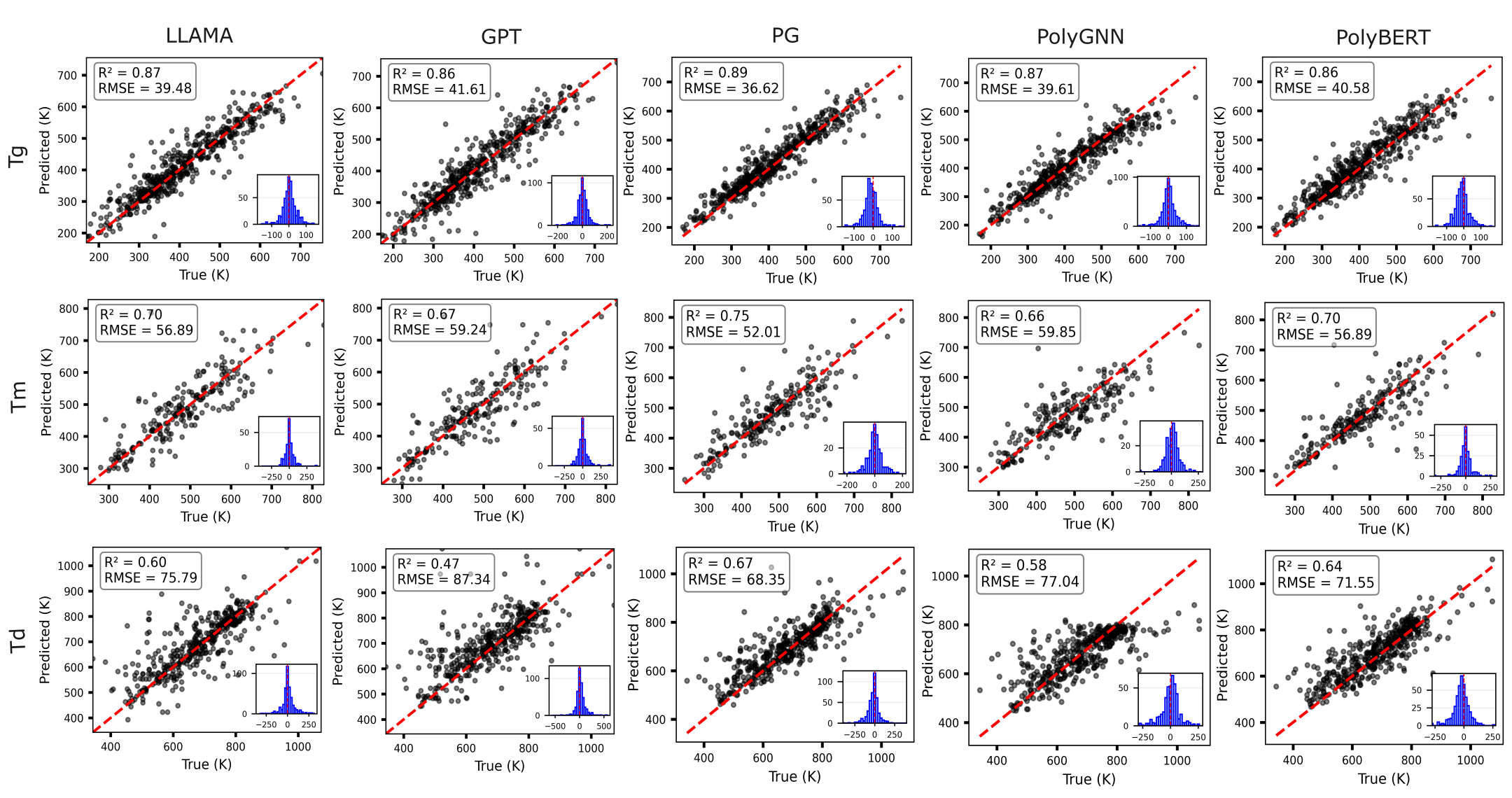}
    \caption{
    Parity plots for T$_g$, T$_m$, T$_d$ under ST learning using the fine-tuned LLaMa-3 model ($\alpha = 16$, $r = 16$, epochs = 25, $T = 0.8$), fine-tuned GPT-3.5 model (epochs = 5, $T = 0.5$), and traditional fingerprinting-based models: PG, polyGNN and polyBERT.
    }
    \label{fig:fig2}
\end{figure*}


\textbf{Fig. 3} presents the comparative model performance under Single-Task (ST) and Multi-Task (MT) configurations. With optimal hyperparameter settings, fine-tuned LLaMA-3 (a=16, r=16, epochs = 25, T = 0.8) consistently outperformed GPT-3.5 (epochs = 5, T = 0.5) across all datasets. The ST models for LLaMa-3 achieved lower RMSE values in predicting \( T_g \), \( T_m \), and \( T_d \), recording 39.48 K, 58.23 K, and 77.11 K, respectively compared to 47.2 K, 63.8 K, and 80.5 K for GPT-3.5. These results indicate the efficiency of the smaller, fine-tuned LLaMA-3 model over GPT-3.5 under single-task learning for thermal property prediction.

\begin{figure*}
    \centering
    \includegraphics[width=3.25in]{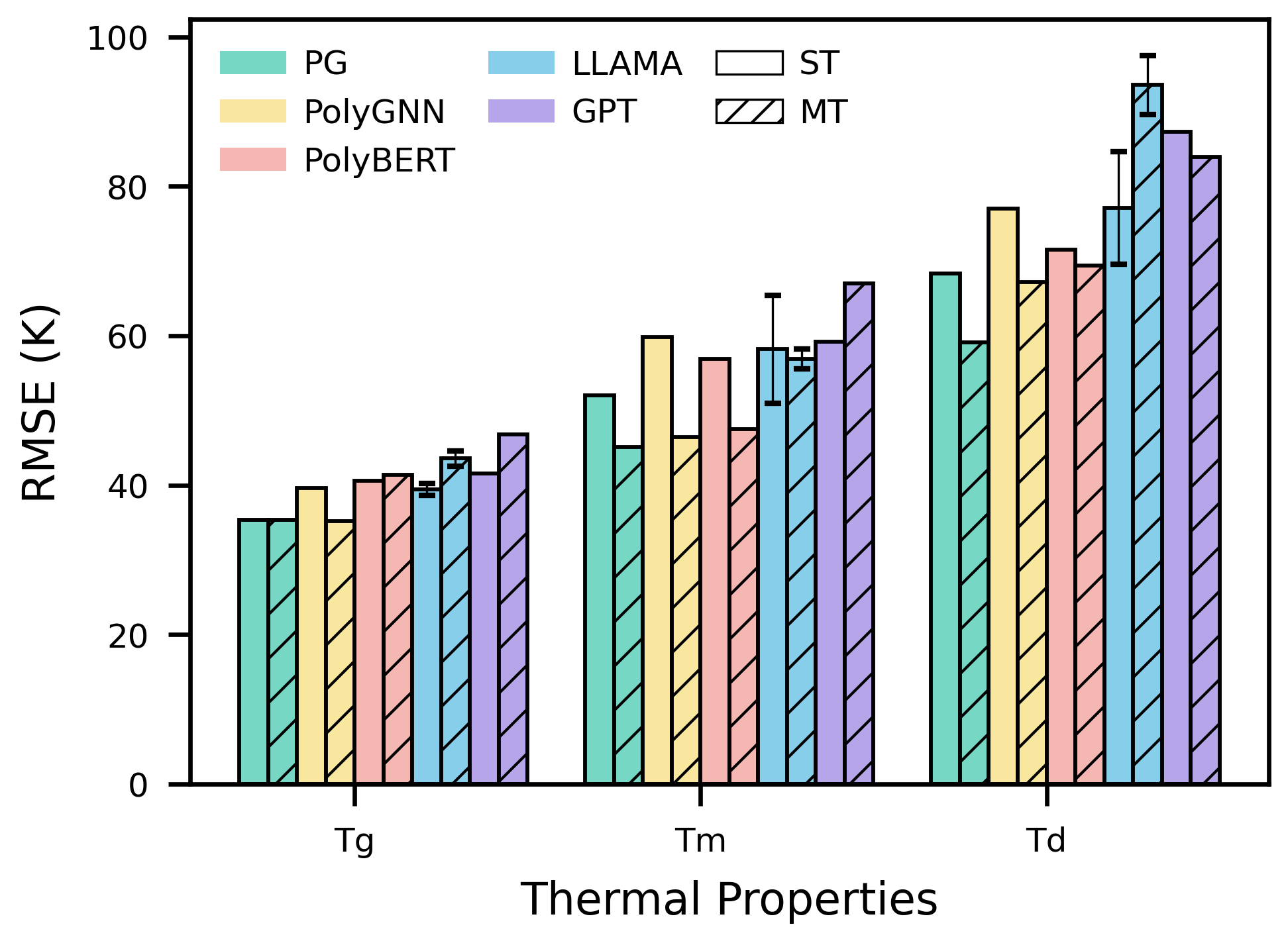}


    \caption{Comparison of \(T_g\), \(T_m\), and \(T_d\) predictions for PG, PolyGNN, PolyBERT, and fine-tuned LLaMA-3-8B-Instruct (LoRA: \(a = 16, r = 16, T = 0.8, \text{epochs} = 25\)) and GPT-3.5 models (\(\text{epochs} = 10, T = 0.5\)) under the single-task (ST) and multi-task (MT) learning framework. Error bars for LLaMA-3 models were calculated based on three separate inference runs for each configuration. GPT-3.5 models do not include error bars, as only a single run was performed for each configuration to reduce costs.}

\end{figure*}

Under the Multi-Task (MT) configuration, neither LLaMA-3 nor GPT-3.5 demonstrated substantial improvements over their ST counterparts. For both \( T_g \) and \( T_m \), the RMSE values were similar to those of the ST models, though performance for \( T_d \) showed a degradation in accuracy. As shown in \textbf{Fig. 3}, PG, PolyGNN, and PolyBERT benefited from MT learning, while LLaMA-3 and GPT-3.5 models failed to achieve similar improvements, suggesting that LLMs struggle to efficiently share learned representations across multiple polymer properties.
This is a notable divergence of the LLMs from the trends observed in traditional ML-based methods, where multi-task learning typically enhances predictive performance by leveraging cross-property correlations \cite{kuenneth_polybert_2023, gurnani_polymer_2023}. 

Interestingly, the fine-tuned LLaMA-3 models consistently achieved lower RMSE values than fine-tuned GPT-3.5 models, highlighting its greater adaptability. While the GPT-3.5 models required fewer epochs and incurred lower on-premise computational costs, its black-box nature and limited hyperparameter tuning options restricted its optimization potential. In contrast, LLaMa's open-source flexibility enabled more effective fine-tuning, which potentially led to a better predictive accuracy.
Ultimately, the advantage of LLMs lies in their scalability and ease of fine-tuning compared to traditional ML models, which require extensive domain-specific customization and computationally expensive pretraining. However, domain-specific approaches still hold an edge in predictive accuracy, emphasizing the need for future research to integrate the strengths of LLMs with specialized polymer informatics models.

\subsection*{Continual Learning and Forgetfulness of LLMs}
To explore the feasibility of continual learning, referred to in this paper as MT-Seq, we tested a method where models were fine-tuned iteratively to study a hypothetical scenario where additional data or data for new properties become available at a later time. The goal was to evaluate whether a model could retain previously learned knowledge while integrating new information. Two tests were conducted to investigate this.

In the first test, sequential training was applied to the $T_g$ dataset only. The dataset was split into 90\% training and 10\% testing data. The training set was further divided into two halves, with the model being fine-tuned iteratively: the first iteration involved fine-tuning on the first half of the training data, and the second iteration fine-tuned the model using the remaining half. The same testing dataset was used to evaluate performance across iterations. As shown in \textbf{Fig. 4a}, the fine-tuned LLaMA-3 model achieved RMSE of 60.76 K after the first iteration and 56.81 K after the second iteration, demonstrating improved performance as additional data was introduced.
Similarly, for GPT-3.5, a similar improvement in performance was noted, where the RMSE slightly decreases from 43.73 K in the first iteration to 42.22 K in the second.

\begin{figure*}
    \centering
    \includegraphics[width=6.48in]{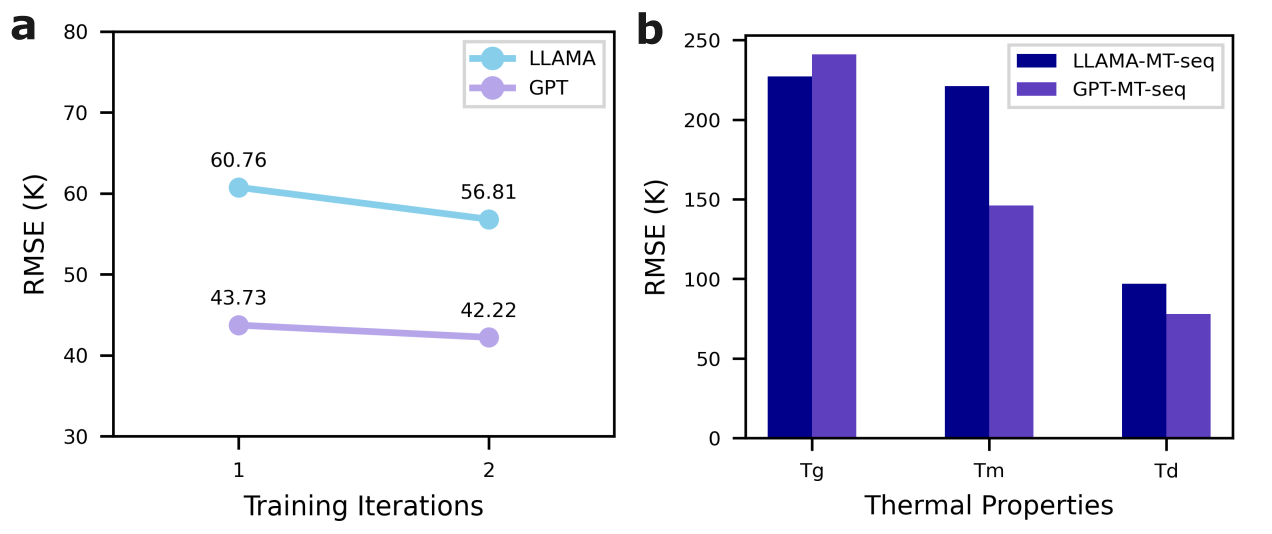}
    \caption{\justifying Performance evaluation of continual fine-tuning experiments. (a) RMSE values for the \(T_g\) dataset across two iterations of fine-tuning using the LLaMA-3 and  GPT-3.5 models. (b) RMSE values for continual fine-tuning on \(T_g\), \(T_m\), and \(T_d\) datasets under the Multi-Task-Sequential (MT-Seq) configuration, comparing LLaMA-3 and GPT-3.5 models.}

\end{figure*}

The second test extended this sequential training approach to the entire thermal properties dataset, where LLMs were first fine-tuned on the (\(T_g\) dataset, next on the \(T_m\) dataset, and finally on the \(T_d\)) dataset. Both GPT-3.5 and LLaMA-3 showed degraded performance compared to the previously discussed ST and MT training configurations. For the GPT-MT-Seq configuration, RMSE values for \(T_g\) and \(T_m\) significantly increased during the later stages of training, indicating that the model failed to retain previously learned information. However, for \(T_d\), GPT’s performance remained comparable to the MT and ST configurations, potentially due to overfitting on the \(T_d\) dataset. Interestingly, GPT-3.5 exhibited an intuitive trend where the oldest dataset (\(T_g\)) suffered the greatest forgetting, while information about more recent dataset, (\(T_m\)) were retained to a greater degree. In contrast, LLaMA-MT-Seq showed consistent performance degradation across all three datasets, without any clear retention of recent data, further highlighting its susceptibility to the phenomenon known in computer science as `catastrophic forgetting'.

These results underscore the limitations of continual learning in LLMs, with both models exhibiting `catastrophic forgetting' when fine-tuned iteratively on datasets. These findings align with existing literature, where continual fine-tuning has shown to result in similar forgetting behavior \cite{luo2024empiricalstudycatastrophicforgetting}. This phenomenon, also known as the palimpsest effect, occurs when newly learned information overwrites previously acquired knowledge \cite{zenke2024theoriessynapticmemoryconsolidation}.

\subsection*{Effects of Polymer Representations via Embeddings}

Leveraging molecular embeddings as input feature vectors is essential to achieve accurate predictions of chemical properties using traditional machine learning \cite{goh2017smiles2vec}. For embeddings to be effective, they must not only capture the intricate chemo-structural attributes of materials but also remain robust to transformations that leave the material's physical state unchanged. Over the years, a variety of approaches- such as, PG , polyGNN and polyBERT fingerprinting- have been developed to generate such robust embeddings \cite{doan_tran_machine-learning_2020, gurnani_polymer_2023, kuenneth_polybert_2023}. 

\begin{figure*}
    \centering
    \includegraphics[width=6.48in]{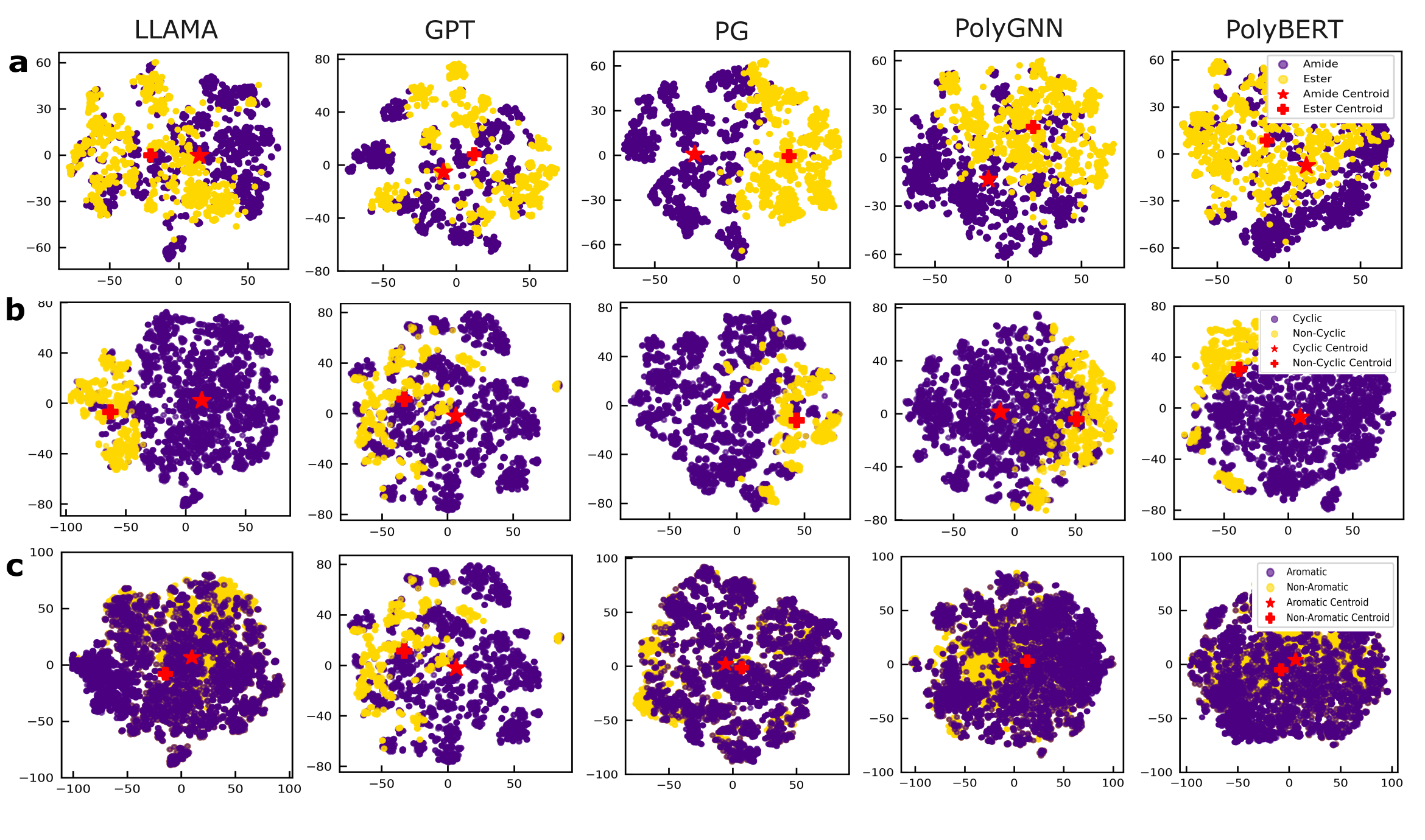}
    \caption{Two-dimensional t-SNE projections of polymer embeddings, showcasing: (a) amide vs. ester functional groups, (b) cyclic vs. non-cyclic polymers, and (c) aromatic vs. non-aromatic polymers.}
    \label{fig:embeddings}
\end{figure*}

To assess the chemical understanding captured by traditional fingerprinting methods and general-purpose LLM embeddings, we compared fingerprints generated using PG, polyGNN, polyBERT with those produced by OpenAI’s text-embedding-3-small model, and LLaMA-3-8B embeddings. For LLaMA-3, the embeddings were extracted from the output of the model's final layer. The quality of these embeddings was assessed based on their ability to differentiate between distinct polymer structural features, including functional groups, cyclic vs non-cyclic structures, and aromatic vs non-aromatic compounds. \textbf{Fig. ~\ref{fig:embeddings}} showcases scatter plots of two-dimensional t-SNE projections of embeddings for three comparisons: (a) amide vs. ester functional groups, (b) cyclic vs. non-cyclic structures, and (c) aromatic vs. non-aromatic compounds. Polymers with amide and ester functional groups were selected for analysis due to their substantial representation in the dataset, with 1,920 and 1,457 data points, respectively, ensuring a fair comparison across methods. Similarly, cyclic and aromatic classes were chosen for their relevance in structural and chemical diversity, providing additional insight into the embedding quality. Centroids for each cluster, calculated as the mean of all data points within a cluster, are also included for clarity.

\textbf{Fig. ~\ref{fig:embeddings}a} illustrates the embeddings derived for polymers with amide and ester functional groups. The PG fingerprints exhibit distinct and well-separated clusters, demonstrating the effectiveness of handcrafted features in capturing the chemo-structural attributes of these polymer classes. PolyGNN and PolyBERT embeddings also show clustering for the two classes, though the separation is less pronounced compared to PG. LLaMA-3 embeddings, while not as refined as the domain-specific methods, still reveal evident clustering, showcasing its capability to capture underlying chemical and structural differences from SMILES representations. In contrast, embeddings generated by GPT display minimal separation between the two functional groups, indicating limited clustering. This suggests that GPT struggles to distinguish between these chemical classes, elucidating a deficiency in its understanding of chemical relationships when compared to both handcrafted fingerprints and domain-specific embeddings.

\textbf{Fig. ~\ref{fig:embeddings}b} showcases embeddings for cyclic vs. non-cyclic polymers, with the PG fingerprinting taken as the reference due to its superior performance across tests. Both PG and LLaMA-3 embeddings demonstrate a clear separation between the two classes, effectively capturing the structural distinctions. In contrast, GPT embeddings show significant overlap and interspersed clusters, highlighting its potentially limited ability to differentiate between cyclic and non-cyclic polymers. Similarly, \textbf{Fig. ~\ref{fig:embeddings}c} compares aromatic and non-aromatic compounds. All models exhibit overlapping clusters in this case, reflecting the inherent difficulty of distinguishing between these two classes in a 2D representation. While LLaMA-3 embeddings provide slightly better separation than GPT, the overlap remains significant. Distance between the centroids of the clusters further emphasizes these overlaps, illustrating the challenges and limitations of these embeddings in effectively capturing chemical distinctions in this context. These observations are consistent with prior studies, where LLaMA-2 demonstrated superior performance in generating molecular embeddings from SMILES strings compared to GPT's `text-small-3-embeddings' embedding model~\cite{sadeghi2024can}.

Overall, these results emphasize LLaMA-3's ability to generate embeddings which better capture chemical and structural distinctions compared to GPT-3.5, particularly for distinct classifications like cyclic vs. non-cyclic polymers. However, the findings also highlight the importance of chemically informed embeddings, such as those generated by PG and domain-specific ML models, in achieving superior representation of polymers.

\section*{Conclusion}
In this study, we investigated the capabilities of LLMs in polymer informatics, benchmarking their performance against traditional domain-specific approaches. The findings demonstrate that while LLMs are effective for property predictions of polymers, their performance is significantly shaped by task-specific requirements, model configurations and the quality of the learned embeddings. Notably, while LLMs can reach the performance of the SOTA ML models, they are not able to surpass them, highlighting the ongoing need for optimization and hybrid approaches.
Notably, ST learning methods consistently outperformed MT approaches using LLMs, a deviation from the trends seen in the traditional ML models. This suggests that LLMs benefit from specialized tuning for individual properties rather than relying on shared parameter spaces across multiple properties.
Another limitation of LLM finetuning is catastrophic forgetting—a significant challenge when integrating new data without compromising previously learned knowledge. These findings highlight the need for customizing training strategies to align with both the model architecture and dataset characteristics.

A comparison between GPT-3.5 and LLaMa-3 revealed that the smaller LLaMa-3 model achieved superior performance, showcasing both its efficiency and adaptability. However, analysis of RMSE parity plots exposed limitations such as clustering and reduced variability in predictions. These findings emphasize the inadequacy of relying solely on traditional metrics like RMSE to evaluate performance. Advanced metrics, such as the Bin Height Dispersion Index (BHDI) introduced here, enabled a deeper and more nuanced assessment of model predictions.

Embeddings emerged as a critical factor in model performance. Handcrafted fingerprints and domain-specific embeddings excelled at capturing chemo-structural details, leading to more accurate property predictions. In contrast, GPT embeddings exhibited insufficient chemical understanding, leading to poor differentiation and clustering of polymer classes. LLaMA-3’s embeddings, on the other hand, demonstrated greater consistency and effectively captured critical aspects of polymer chemistry. This advantage, coupled with its flexibility in hyperparameter tuning, likely contributed to LLaMA-3's superior performance over GPT-3.5. 

Overall, while LLMs have demonstrated substantial promise in polymer informatics, their current performance suggests they provide ease rather than accuracy relative to SoTA traditional models, atleast for the type of tests considered here. Future research should focus on integrating domain knowledge with LLM-driven approaches, refining training methodologies, and improving embedding strategies to enhance their generalizability and predictive power in polymer property prediction.

\section*{Methods}\label{sec:methods}

\subparagraph{Prompt Optimization}

To evaluate the impact of prompt design on model performance, we conducted prompt optimization experiments using GPT-3.5 on a dataset for $T_g$. Given the cost constraints associated with using GPT-3.5, a 10/90 data split was employed, with only $10\%$ of the dataset used for training and $90\%$ for testing. Inference was performed with a temperature (T) of 0.5. These considerations were made to maximize insights while minimizing computational expenses.
Five different prompt designs were tested, each varying in structure and the specificity of the query-response format. These prompts were carefully crafted to assess their influence on the model's ability to generate accurate predictions, as measured by Root Mean Squared Error (RMSE). 

The results, as shown in Fig S5., indicate significant variability in RMSE across the prompts. Prompt 01, which employed a straightforward and well-structured question-response format, achieved the lowest RMSE, highlighting its effectiveness in eliciting accurate predictions. In contrast, Prompt 04, designed to handle uncertainty, resulted in the highest RMSE, likely due to the added complexity and potential for generating ambiguous responses. Similarly, Prompt 03 showed a relatively high RMSE, which may be attributed to its overly simplistic format that asked the model to predict only a numeric value. These findings emphasize the importance of prompt design in optimizing the predictive capabilities of LLMs. A well-balanced prompt structure, as demonstrated by Prompt 01, structured as:  
\textbf{User}: \texttt{If the SMILES of a polymer is <SMILES>, what is its <property>?}  \textbf{Assistant}: \texttt{\{smiles: <SMILES>, <property>: <value> <unit>\}} enables the model to leverage its contextual understanding effectively while maintaining accuracy.
 
\subparagraph{LLaMA Model}
The Meta LLaMA-3-8B-Instruct model was employed for fine-tuning, leveraging pretrained weights available on the Hugging Face hub via the Transformers library. To enhance computational and memory efficiency parameter efficient fine-tuning was performed with Low-Rank Adaptation (LoRA) \cite{hu2021loralowrankadaptationlarge} method on two NVIDIA L40S GPUs (46 GB, 350 W) hosted on our in-house servers. Hyperparameter optimization included variations in alpha and rank values (16, 32, 64, and 128), with the number of epochs tested incrementally from 5 to 30. Inference temperatures of 0.5 and 0.8 were also explored to achieve optimal performance.

\subparagraph{GPT Model}
The GPT-3.5-turbo-0125 model was fine-tuned through the OpenAI API, accessed via the OpenAI Python package. Although GPT-4o-2024-08-06 was initially tested, its performance was comparable to GPT-3.5, leading to the selection of the latter due to its cost efficiency. Performance comparisons between the two models are detailed in the Supporting Information. The use of an API-based approach streamlines deployment but limits flexibility in hyperparameter control. Optimization focused on the number of epochs (varied between 5 and 10) and inference temperatures (evaluated at 0.5 and 0.8).

\subparagraph{Evaluation Metrics.}\label{sec:evaluation_metrics}
Model performance was evaluated using the coefficient of determination ($R^2$) and the root mean square error ($RMSE$), standard metrics in materials science for assessing precision and the fit between predictions and ground truth. However, these metrics fail to capture the distributional characteristics and clustering of the predictions. To address the gap, we introduce Bin Height Dispersion Index (BHDI), a complementary metric designed to assess the uniformity and variability of model predictions. BHDI quantifies the clustering of predicted values, often observed as horizontal lines in parity plots or sharp spikes in Predicted vs. Ground Truth distribution plots, offering deeper insights into prediction diversity and enhancing the overall evaluation of model performance.


\begin{equation}
R^2 = 1 - \frac{\sum_{i=1}^{n} (y_i - \hat{y}_i)^2}{\sum_{i=1}^{n} (y_i - \bar{y})^2}
\end{equation}

\begin{equation}
RMSE = \sqrt{\frac{1}{n} \sum_{i=1}^{n} (y_i - \hat{y}_i)^2}
\end{equation}

\subparagraph{Bin Height Dispersion Index.}
Bin Height Dispersion Index (BHDI) quantifies clustering by measuring the deviation of bin heights in the histogram $h_{pred}$ of the predicted values from those in the histogram $h_{true}$ of the ground truth values. We define BHDI as:

$$
     BHDI = \frac{\sum_{i=1}^{N}(h_{pred}^i - h_{true}^{i})^2}{\sum_{i=1}^{N} h_{true}^i}
$$

where, \(N\) represents the total number of bins in the histograms, \(h_{\text{true},i}\) is the height (number of times a property value is predicted) of the \(i\)-th bin in the ground truth histogram, and \(h_{\text{pred},i}\) is the corresponding height in the predicted histogram. The denominator, \(\sum_{i=1}^{N} h_{\text{true},i}\), represents the total frequency count of the ground truth histogram, ensuring normalization. 
A lower BHDI value reflects a well-distributed range of predictions, indicating good variability and robust generalization across the data space. In contrast, a higher BHDI value highlights significant clustering, where predictions are overly concentrated in specific bins, signaling poor generalization and limited variability. By addressing the limitations of traditional evaluation metrics, BHDI offers a more comprehensive assessment of model performance, especially in tasks like property prediction where balanced and diverse predictions are critical.


\section*{Data Availability}
Data sharing is not applicable to this article as no new data was created or analyzed in this study.

\section*{Acknowledgement}
This work was supported by the Office of Naval Research through grants N00014-19-1-2103 and N00014-20-1-2175.

\section*{Competing Interests}
The authors declare no competing interests.

\section*{Supporting Information}
The online version of this article contains Supplementary Information available at \url{https://doi.org/}
\begin{itemize}
    \item Prompt optimization, parity plots for Llama and GPT finetuned models etc.
\end{itemize}

\def\bibsection{\section*{References}}
\bibliography{2024-LLMInformatics}
\end{document}